\documentclass[runningheads]{llncs}
\usepackage{graphicx}
\usepackage{cite}
\usepackage{amsmath,amssymb,amsfonts}
\usepackage{algorithmic}
\usepackage{graphicx}
\usepackage{adjustbox}
\usepackage{textcomp}
\usepackage{xcolor}
\def\BibTeX{{\rm B\kern-.05em{\sc i\kern-.025em b}\kern-.08em
    T\kern-.1667em\lower.7ex\hbox{E}\kern-.125emX}}
    
\usepackage{caption}

\usepackage{subcaption}
\usepackage[linesnumbered,ruled,vlined]{algorithm2e}

\SetCommentSty{mycommfont}

\SetKwInput{KwInput}{Input}                
\SetKwInput{KwOutput}{Output} 

\usepackage{hyperref}

\begin{document}
\title{NODDLE: Node2vec based deep learning model for link prediction
\thanks{This research is funded by NSERC Discovery Grant (RGPIN-2017-05377), held by Dr. Vijay Mago.}
}
%
%

%
%
%
%
\author{Kazi Zainab Khanam\orcidID{0000-0003-1106-4598} \and
Aditya Singhal\orcidID{0000-0001-9634-4075} \and
Vijay Mago\orcidID{0000-0002-9741-3463}}

\authorrunning{K. Khanam et al.}

\institute{Lakehead University, Thunder Bay ON P7B 5E1, Canada \\
\email{\{kkhanam,asinghal,vmago\}@lakeheadu.ca}}

\maketitle              
\begin{abstract}
Computing the probability of an edge’s existence in a graph network is known as link prediction. While traditional methods calculate the similarity between two given nodes in a static network, recent research has focused on evaluating networks that evolve dynamically. Although deep learning techniques and network representation learning algorithms, such as node2vec, show remarkable improvements in prediction accuracy, the Stochastic Gradient Descent (SGD) method of node2vec tends to fall into a mediocre local optimum value due to a shortage of prior network information, resulting in failure to capture the global structure of the network. To tackle this problem, we propose NODDLE (integration of NOde2vec anD Deep Learning mEthod), a deep learning model which incorporates the features extracted by node2vec and feeds them into a four layer hidden neural network. NODDLE takes advantage of adaptive learning optimizers such as Adam, Adamax, Adadelta, and Adagrad to improve the performance of link prediction. Experimental results show that this method yields better results than the traditional methods on various social network datasets.

\end{abstract}
\section{Introduction}
Link prediction is a fundamental problem of network analysis, mainly because of its importance in social network applications such as designing recommendation systems for social media platforms and e-commerce websites, identification of credit card fraud, and even locating terrorist groups based on their criminal activities \cite{aiello2012friendship,liben2007link,wang2015link, bressan2017counting, chen2017harp,pezeshkpour2019investigating}. The field of bioinformatics often uses link prediction for predicting protein-protein interactions containing important information about biomolecular behavior. Such interactions can reveal answers about diseases and cures \cite{airoldi2008mixed}, and therefore, predicting such upcoming links is a crucial component of graph mining.

The main objective of the link prediction problem is to predict the unseen edges that will emerge in a graph. Based upon the \textit{snapshot assumption}, when a snapshot of a graph \textit{G(t)} at time \textit{t} is given, link prediction is used to compute which new upcoming links will emerge in the future graph $G(t^\prime)$ within the time period $[t,t^\prime]$, where $t^\prime = t + n$ ($n$ is the sequence of snapshots) \cite{liu2019link}. Link prediction is implemented on real-world network graphs, which are often too massive and \textit{dynamic}, as such graphs are evolving at an extremely high speed. In addition, link prediction uses proximity-based measures, such as the Jaccard coefficient, Resource Allocation, and Adamic Adar metric, to measure the probability of the upcoming links in the network \cite{ayoub2020accurate}. The features extracted are based on the local nodal properties as these functions use the information available only from the local proximity of the nodes. Although these metrics are used widely in multiple applications because of their simplicity and interpretability, the problem arises when social network graphs become large with numerous users. As a result, predicting future links with these measures becomes a very challenging task. Most importantly, hidden and meaningful knowledge lies between the nodes and edges of networks \cite{airoldi2008mixed}, and analyzing these graphs is extremely difficult when large-scale network data comprises billions of nodes and edges \cite{tang2015line}.

Traditional approaches calculated link prediction statically by using only a single network snapshot to predict future links. However, the prediction task is a time-dependent problem, where a network evolves over time \cite{rahman2016link}. Hence, the dynamic network concept was initiated in which the structure of the network is captured in multiple snapshots over a span of time \cite{zhang2017efficient}. Dynamic-link prediction is considered more valuable and challenging than static link prediction. The evolvement of the network structure offers much more information that adds a whole new dimension in network analysis and helps achieve a better link prediction performance \cite{rahman2018dylink2vec}. The problem arises when the number of edges and nodes increases at a faster rate as it becomes very challenging to extract or infer any reasoning and information from the whole network \cite{grover2016node2vec}. Dimensional reduction techniques have been used to solve this issue, which transform the nodes of a graph into lower-dimensional latent representations \cite{al2019ddgk}. These representations can be used as features for executing tasks in graph mining, such as clustering and link prediction \cite{xu2018interaction}. Similarly, network representation learning algorithms such as node2vec have also been used to tackle this issue. Node2vec conducts high order proximity by escalating the probability of finding successive neighboring nodes within a fixed length of random walk \cite{grover2016node2vec}. This method can efficiently find the equilibrium position between breadth-first search (BFS) and depth-first search (DFS) graphs by developing random biased walks. As a result, it can succeed in embedding rich quality data, enabling node2vec to preserve the structural balance of the node communities. 

Although node2vec has successfully achieved high link prediction performance, it still has many shortcomings \cite{chen2019n2vscdnnr}. Firstly, it follows a local approach that takes short random walks to get exposed to only the local neighborhood of nodes \cite{chen2018harp} and hence ignores the global relationship of nodes that might have longer distances. Due to this, the learned representation may be unable to comprehend the essential global structure of the model. Secondly, node2vec uses the Stochastic Gradient Descent (SGD) method to resolve non-convex optimization problems, where the non-convex constraints may have various regions and many locally optimal points within each region \cite{amari1993backpropagation,jain2017non}. The algorithm repeatedly gets updated when SGD is used to optimize the objective function. This causes the optimal points to oscillate frequently and possibly causes them to get stuck in a local minima. Due to the complexity of the growing networks, recent research has focused on applying deep learning techniques to evaluate the complex relationships that exist in graphs and visualize the hidden patterns \cite{wang2017relational}. To tackle these problems we propose NODDLE (integration of \textbf{NO}de2vec an\textbf{D} \textbf{D}eep \textbf{L}earning m\textbf{E}thod), a deep learning model that combines the features extracted by node2vec algorithm and feeds them into four layers of hidden neural network. It optimizes the performance by using different types of optimizers, which include Adaptive Moment Estimation (Adam), Adamax, An Adaptive Learning Rate Method (Adadelta), and Adaptive Gradient Algorithm (Adagrad). We have compared our approach with the benchmark methods that include Adamic Adar, Jaccard coefficient, and Preferential Attachment \cite{duchi2011adaptive,kingma2014adam,zeiler2012adadelta}.

The rest of the paper is organized as follows. Section \ref{Related Work} presents background on the previous studies conducted on link prediction of social networks with heuristic-based, machine learning, and deep learning approaches. Section \ref{Proposed} introduces our proposed approach in detail, including our data preparation method and the method for combining node2vec with the deep learning model. Later, in section \ref{Experiments}, we validate our approach on real-world social network data and analyze the
results. Finally, Section \ref{Conclusion} concludes the paper.

\section{Related Work} \label{Related Work}

\subsection{Heuristic Similarity Metrics}

Liben-Nowell and Kleinberg proposed a link prediction problem for social networks using multiple heuristic functions \cite{liben2007link}. They found that topological features can be used to predict a future edge between two nodes that showed high ``similarity'' or ``proximity'' between the target nodes. Furthermore, their findings conveyed that the heuristics such as Adamic/Adar and Katz centrality measure notable correlation with the predicted future links \cite{adamic2003friends, katz1953new}.

Many research works emphasized enhancing the performance of heuristic functions by increasing the neighbor-based attributes to second, third, or higher adjacency degrees. For instance, Yao et al. presented an improved common neighbors heuristic algorithm that includes nodes with a distance of two hops and used time-decay for recent snapshots to have a greater weight \cite{yao2016link}. Kaya et al. used progressive events to calculate the possibility of future links in a time-weighted fashion \cite{kaya2017unsupervised}, and Deylami and Asadpour proposed a community detection algorithm to identify high activity clusters \cite{deylami2015link}. Similarity metrics have also been used to detect social and cognitive radio network events for common link prediction problems \cite{hu2017retracted,hu2017event,zhang2016analytical}.

\subsection{Machine Learning \& Deep Learning}

Both supervised and unsupervised techniques have been employed to predict links in the network. Unsupervised methods comprise developing the heuristic approaches to determine the score for the likelihood of each upcoming link \cite{tabakhi2014unsupervised}. Similarity metrics are most commonly used to measure the intensity of the relationship between the nodes. Topological features of the nodes such as common neighbors and graph distances are used to measure the strength of the interaction between the nodes \cite{al2006link}. Conversely, supervised methods involve treating the link prediction problem as a binary classification task in which the edges and non-edges of a network model are employed for training a classifier \cite{li2018link}.

Compared to heuristic-based approaches, machine learning techniques have proven better at link prediction tasks as these models have received higher accuracy. Yet, the major problem with them is representing the graphical features, since it is impossible to use the large-scale graphs as input into the machine learning models. As a result, researchers have attempted to extract features. For example, Hasan et al. have extracted multiple graph features and implemented the features with various machine learning algorithms such as Decision trees, Naive Bayes, and k-Nearest Neighbors \cite{al2006link}. Similarly, Bechettara et al. have implemented topological-based features of bipartite graphs with decision trees \cite{benchettara2010supervised}, and Doppa et al. proposed a supervised feature vector-based approach with k-means classifier for link prediction \cite{doppa2010learning}. Even though machine learning techniques have been shown to achieve better prediction accuracy, these methods rely highly on features developed by human intelligence. Thus, engineering such features is extremely tedious and slow. As a result, most state-of-art link prediction techniques utilize deep neural networks for their exceptional learning ability.

A deep neural network model is defined as a group of models in machine learning consisting of multiple connected layers. The layers generate output-yielding nodes where the parameters of the neural network layers are tuned in continuous iterations to reduce the error between the final output and the original value \cite{canziani2016analysis}. Li et al. have explored a neural network structure as a conditional temporal Restricted Boltzmann Machine (ctRBM), which expands on the architecture of an RBM to integrate the temporal elements of a dynamic changing network \cite{li2014deep}. Furthermore, Zhang et al. suggested the neural network model as a means of feature representation by using the term Social Pattern and External Attribute Knowledge (SPEAK); these features are used as input in deep neural network models \cite{zhang2016deep}. Ozcan A has proposed a link prediction algorithm that extracts multi-variable features from heterogeneous networks and is based upon non-linear autoregressive neural networks \cite{ozcan2018link}. This method was tested on various datasets and has outperformed the existing algorithms that focus on only single variable features. Zhang et al. proposed a framework that uses graph neural networks to learn general graph features for link prediction \cite{zhang2016deep}. Graph neural networks are defined as a message-passing algorithm, in which the message represents the features extracted from each node in a graph, and their effects on the edges and nodes are learned by neural networks \cite{maron2019provably}. Their framework has also shown promising results in the online social networking Stanford Facebook dataset \cite{zhang2016deep}. Therefore, state-of-the-art research has mainly focused on learning multiple features from graphs at an extensive level as such features contain hidden and meaningful insights into link probability. With the rise of complex growing networks, deep learning techniques have produced highly accurate results. Besides, deep learning can model the complex relationships hidden in the network data and can reveal unseen patterns hidden beneath the billions of nodes and edges \cite{rahman2016link}.

Further research is being conducted to improve link prediction performance by applying both supervised, unsupervised, as semi-supervised approaches \cite{liu2020network,zhu2020semi}. Semi-supervised learning combines a small proportion of labeled data with a large pile of unlabeled data during the training process. As mentioned earlier, a semi-supervised approach such as node2vec has outperformed existing supervised approaches since it can maintain the community structure and embed better quality information \cite{grover2016node2vec}. In addition, neural networks are also currently being used to enhance link prediction performance. These novel methods have proven to be highly effective \cite{li2018deep}. Such methods can produce promising link prediction results in large complex networks. Even so, a primary disadvantage to such approaches is that the training and prediction process is highly time-consuming.

\section{Proposed Approach}\label{Proposed}

\begin{figure}[htp]
    \centering
    \includegraphics[height=8cm]{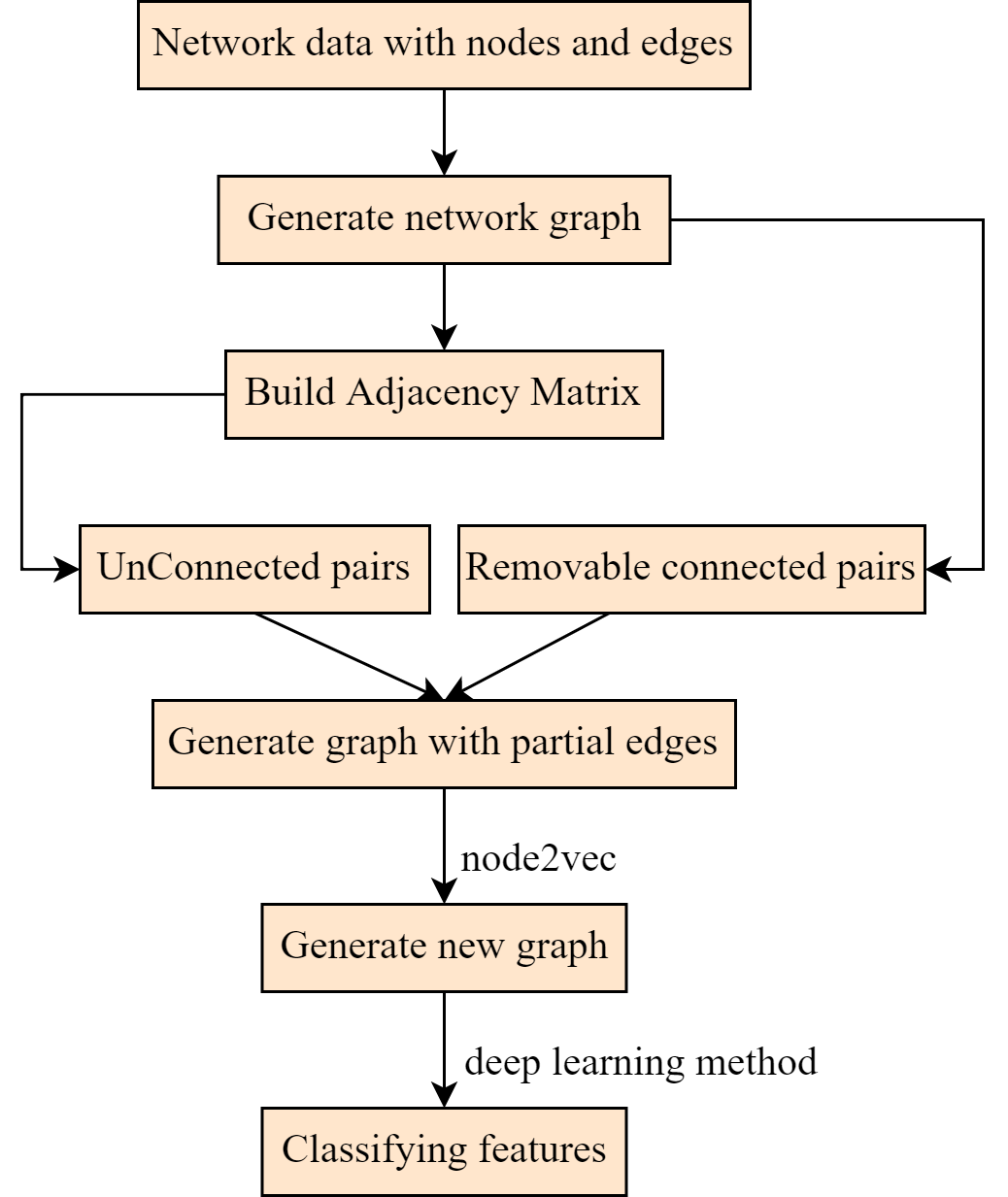}
    \caption{Overall structure of the proposed approach}
    \label{fig:1}
\end{figure}

This section explains the strategy of solving the problem with the deep learning method. Algorithms \ref{Unconnected} and \ref{Connected} provide insights into how we have aggregated the connected and unconnected pairs from the network used to build the training dataset. The overall steps for preparing a graph with connected and unconnected pairs from the raw network graph are explained in Algorithm \ref{DataPrep}. Then, we present the node2vec model for extracting the features from the training network dataset. Finally, we show how we have developed the deep neural network model with improved optimizers for executing AUC scores for the link prediction of the network. Fig.~\ref{fig:1} provides the overview of our proposed approach.

 \vspace{-10mm}
\subsection{Problem Statement}
\begin{figure}
\centering
\begin{minipage}{.5\textwidth}
  \centering
  \includegraphics[width=.6\linewidth]{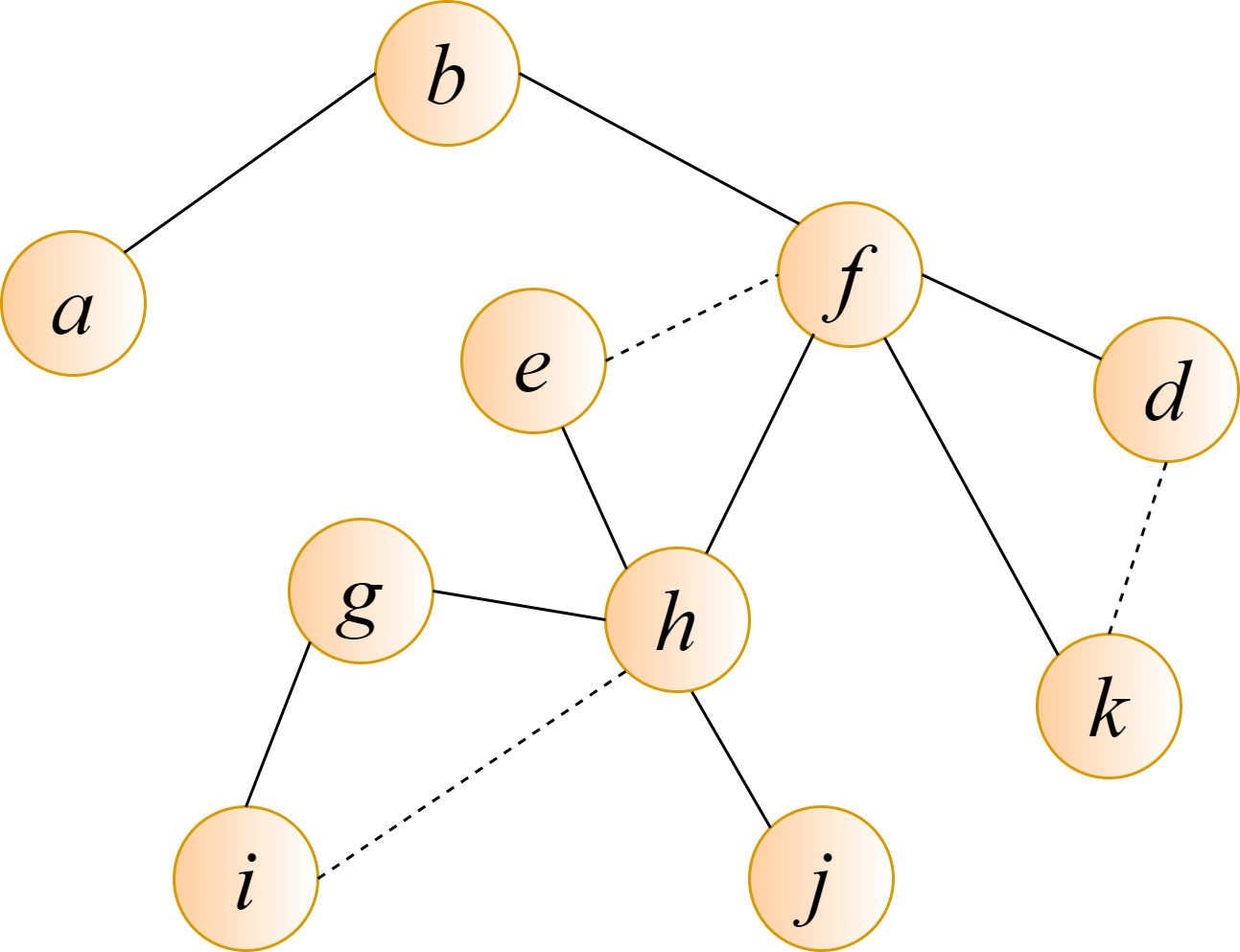}
  \captionof{figure}{At time \textit{t}, the dotted lines represents the future links}
  \label{fig:2}
\end{minipage}%
\begin{minipage}{.5\textwidth}
  \centering
  \includegraphics[width=.6\linewidth]{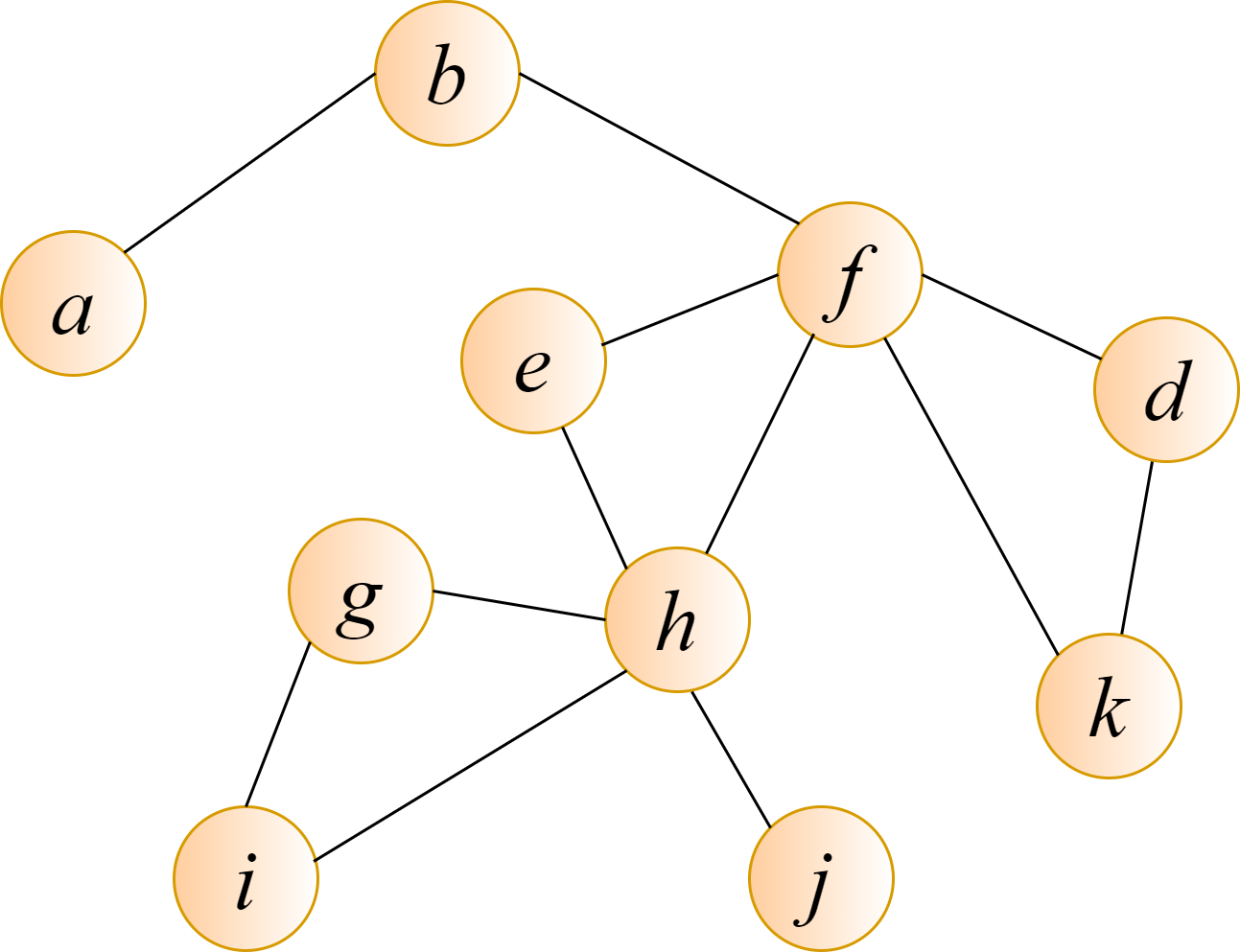}
  \captionof{figure}{At time \textit{t+n}, the new links have formed}
  \label{fig:3}
\end{minipage}
\end{figure}

A sequence of snapshots in time from $t$ to $t+n$ is defined as a dynamic network in which the set of edges in each snapshot depicts the links present at time $t$. The link prediction problem is that given snapshots from $t$ to $t+n$, return the score for the possibilities of edges at time $t+n$. Fig.\ref{fig:2} and Fig.\ref{fig:3} show a dynamic network with two snapshots. Given the information at time $t$, we would like to predict the likelihood of link prediction at time $t+n$.

    

    
\subsection{Data Preparation}

In real-time scenarios, network data is extremely large and highly imbalanced as it contains a higher number of unconnected nodes than connected nodes. Therefore, it is always challenging for the model to learn features from connected nodes since the connected node pairs are often much fewer than the unconnected nodes. Hence, we provide a way for preparing the data computationally economically to extract the unconnected and connected node pairs from a large number of imbalanced network data. Below we discuss the process of sampling the positive (connected) and the negative (unconnected) node pairs.

\subsubsection{Aggregation of Unconnected Samples.}
To find the negative sample that depicts the unconnected nodes, we build an adjacency matrix with the aid of the {\it networkx} library. The connected and unconnected nodes are represented as rows and columns for each node. As the values in the matrix are the same for above and below the diagonal of the adjacency matrix, we only focus on finding the positions of the unconnected nodes from above the diagonal to make the approach computationally efficient. Algorithm \ref{Unconnected} shows the complete steps for finding the unconnected nodes. After experimentation with other configurations, we use {\it networkx} library to find the shortest path between the unconnected nodes and only select the ones within distance 3. The unconnected node pairs were labeled as ‘0', as the unconnected node pairs represent the negative links.

\subsubsection{Aggregation of Connected Samples.}
Some of the edges from the graph will be randomly removed and labelled ‘1’ since, these edges connect the nodes and show the presence of links. Thus, when training the model, it will predict such potential links at time $t+n$. However, it is essential to ensure that the graph's nodes do not become completely isolated when dropping the edges since taking such a step can misrepresent data, and the model will be trained poorly. Thus, when removing an edge, we ensure this does not lead to splitting the graph, and the number of connected nodes is $\geq 1 $ . If the removed edge satisfies both of the conditions, only then the edge is dropped, and the process is repeated for the next pairs of nodes. Algorithm \ref{Connected} shows the steps for accumulating the positive samples.

\begin{algorithm}[h]
\DontPrintSemicolon
  
\KwInput{Adjacency Matrix as $AdjM$, Graph $G$ = $(N, E)$ }
\KwOutput{All unconnected pairs as $UCP$}
$UCP$ $\gets \phi$ \tcp*{empty dataframe for all unconnected pairs}
\For{each row in AdjM}{ 
                \For {each column in AdjM}{
                    \If{\textit{row.index} $\neq$ \textit{column.index}}{
                    \If {FindShortestLength \textit{(G, row, column)} $\leq$ 3}{ \tcc{using networkx for shortest length function}
                    \If {AdjM $[row,column]$ == 0}{
                    Append \textit{N.row}, \textit{N.column} to $UCP$
                    }}}
                }
}
\textbf{Return} UCP
\caption{Finding all Unconnected pairs}
\label{Unconnected}
\end{algorithm}

\begin{algorithm}[h]
\DontPrintSemicolon
  
\KwInput{Graph $G$ = $(N, E)$ }
\KwOutput{Connected Pairs as  $CP$}
CP $\gets \phi$ \tcp*{Empty set of connected pairs}
\For{each e in E}{ 
            $G'\gets$  $RemoveEdge(G)$
        	\tcp{remove edge from a node pair and generate a new graph as $G'$}
        	\If {the new nodes are not completely isolated}{
        	$CP\gets $ Append($G'$.node, $G'$.edge)
            }}

\textbf{Return} $CP$
\caption{Finding all Connected pairs}
\label{Connected}
\end{algorithm}

\begin{algorithm}[h]
\DontPrintSemicolon
  
\KwInput{Node as $N$ and Edge as $E$}
\KwOutput{New Graph as $G'$}
G $\gets \phi$ \tcp*{Graph}
AdjM $\gets \phi$ \tcp*{Adjacency Matrix}
UCP $\gets \phi$ \tcp*{Empty set of unconnected pairs}
CP $\gets \phi$ \tcp*{Empty set of connected pairs}

\For{each n, e in N, E}{ 
        	G  $\gets (n, e)$ \tcp*{Creating network Graph with networkx library}
        	AdjM $\gets$ AdjacencyMatrix $(G) $ \tcc{Create Adjacency Matrix from nodes and edges with networkx}
        	UCP $\gets$ UnconnectedPairs $(AdjM, G) $\tcp*{Algorithm 1}
        	CP $\gets$ ConnectedPairs $(G) $\tcp*{Algorithm 2}
        	$G'$ $\gets $Create NewGraph \textit{(UCP, CP)}
}
\textbf{Return} df
\caption{Data Preparation for generating Model}
\label{DataPrep}
\end{algorithm}

\subsection{NODDLE (integration of NOde2vec anD Deep Learning mEthod)} 
Given that node2vec is a local approach that is limited to the structure around the node, it uses a short random walk to find the local neighborhood of nodes. Such attention to local structure implicitly overlooks the long-distance relationship in the whole network, and the representation may not reveal the important global structure model \cite{grover2016node2vec}. We propose NODDLE, a deep learning model in which features extracted from the node2vec are fed into a 4 layer hidden neural network. The prediction performance is enhanced by applying various optimizers, including Adaptive Moment Estimation (Adam), Adamax, An Adaptive Learning Rate Method (Adadelta), and Adaptive Gradient Algorithm (Adagrad), respectively. We illustrate a brief description of node2vec, its limitations, and explain the need for this deep learning model.

\subsubsection{Node2vec.} 
Node2vec algorithm is a feature extraction method used to generate vector representations of nodes on a graph. It is mainly a local approach that uses random walk to search for the local neighborhood of nodes. The algorithm uses direct encoding and a product-based decoder. Therefore, node2vec embedding is defined as such:  
\begin{equation}\label{twentyt_equ}
DE(s_i,s_j) \cong \frac{e^{z_i^Tz_j}}{\sum_{v_k\in V} e^{z_i^Tz_k}} \approx (P,R (v_j | v_i))
\end{equation}

In this Equation (\ref{twentyt_equ}), $DE (s_i, s_j)$ represents the decoded product based proximity value, the probability of visiting to node target node $v_j$ from the source node $v_i$ with fixed length of random walk $R$, denoted by $(P,R (v_j | v_i))$. $(P,R (v_j | v_i))$ can be calculated for both random and undirected graphs. Cross entropy loss for node2vec is calculated by the following formula: 
\begin{equation}\label{twe_equ}
Loss=\sum_{(v_i,v_j) \in Deno} - \log(DE(s_i,s_j))
\end{equation}

The training set is generated by collecting random walks from a source node $v_i$ in which the $N$ pairs of $v_i$ for each node are collected from the probabilistic distribution of  $(v_i,v_j) \sim (P,R(v_j|v_i))$. However, it is extremely expensive to calculate the cross entropy loss because of the high computational costs for evaluating $O (|Deno| |V|)$, as  $O(|V|)$ has a high time complexity when computing the denominator $Deno$ of (\ref{twentyt_equ}). As a result, node2vec uses various optimization and approximation methods for computing the cross entropy loss. It uses the ``Negative sampling'' approximation method to evaluate (\ref{twe_equ}). The node2vec takes a random set of negative samples for approximately calculating the normalization factor instead of letting the entire set of vertices be normalized \cite{grover2016node2vec}. Additionally, it applies two hyper parameters $p$ and $q$. The probability of going back to a previous node after visiting a new node is controlled by $p$. The hyper parameter $q$ controls the possibility of exploring the graph's new nodes. When these hyper parameters are employed, node2vec can interpolate between the walks much more smoothly, and the approach becomes similar to BFS and DFS. Grover et al. also demonstrated that when the two hyper parameters are well-adjusted, it enables node2vec to preserve the structural balance between the nodes \cite{grover2016node2vec}. However, node2vec still has its drawbacks. It uses SGD method for solving the  non-convex optimization problem \cite{le2014distributed, goldberg2014word2vec}. The algorithm constantly updates when SGD is used as the objective function, which causes the optimal points to oscillate frequently, leading the optimal points to dismount into the local minimum range. 

Besides, SGD keeps the learning rate constant when the parameters are updated. As a result, SGD cannot adapt the learning rate and adjust it for carrying out greater updates on lower frequency features \cite{ruder2016overview}. Hence, Adam, Adamax, and Adadelta optimizers have been introduced to resolve this issue. These optimizers can incorporate different learning rates with different parameters. Compared to SGD, these optimizers are more compatible for large network datatsets in high dimensional spaces and, most importantly, non-convex optimization objective functions. Furthermore, deep learning techniques are also applied to study the complex relationships with the growing networks. Hence, we are focusing on improving the performance of link prediction by fusing node2vec with a deep learning model, in which the model is supported with improved optimizers.

Algorithms \ref{alg:the_alg} and \ref{alg:the_al}, show the steps of the node2vec algorithm. The algorithm first learns the representations of the nodes by generating a random walk with a length of $l$, which starts from each of the nodes. When the step is taken in each of the walks, sampling is conducted with the transitional probability of $\theta_{vx}$. The transitional probability $\theta_{vx}$ of the second order Markov chain is at first calculated so that node sampling can be computed efficiently by using the alias method in \textit{O}(l) time. In the final phase, the transitional probability preprocessing is conducted sequentially, and optimization of SGD is used.

\vspace{-5mm}
\begin{algorithm}
	\caption{Node2Vec Algorithm}
	\label{alg:the_alg}
	\KwInput{Graph $G'$ = ($N, E$), dimension $dim$, Walks per node $r$, Walk Length $l$, Context size \textit{h, Return p, In-out q}} 
    \KwOutput{final Stochastic gradient descent function as  $f$}
	$\theta$ = PreprocessModifiedWeights($G,p,q$)
	$G^{\prime} = (V,E, \theta)$
	Initialize \textit{walks} to empty\\
	\For{iter 1 to r}{	    
	\For{all nodes $ u \in V$}{
	         \textit{walk} = node2vecWalk $(G\prime, u, l)$ \\
	        Append \textit{walk to walks}}}
	 \textit{f} = Stochastic gradient descent (\textit{h, dim, walks})\\
	 \textbf{Return} $f$
\end{algorithm}
\vspace{-15mm}
\begin{algorithm}
    \caption{node2vecWalk Algorithm}
    \label{alg:the_al}
    \KwInput{Graph $G'$ = ($N, E$), Start node $u$, Length $l$, Walk Length $l$, Context size \textit{h, Return p, In-out q}}
    \KwOutput{$walk$}
	 \textbf{node2vecWalk}(Graph $G'$ = ($N, E$), Start node $u$, Length $l$, Walk Length $l$, Context size $h$, Return $p$, In-out $q$) 
	 Initialize $walk$ to [$u$] \\
	 $G' = (V,E,\theta)$ \\
	 Initialize $walks$ to empty\\
	 \For {walk from $1$ to $l$}{
		$curr = walk[-1]$ \\
		 $V_{current}$ = GetNeighbours($Current, G'$) \\
		 $s$ = AliasSample $(V_{current}, \theta)$ \\
	     Append $s$ to $walk$ }
	    \textbf{Return} $walk$

\end{algorithm}
\vspace{-10mm}
\subsubsection{Deep learning Model.}
In the final step, we build a deep learning network in which the features extracted from node2vec are fed into a four layer hidden neural network. As mentioned earlier, the SGD optimization function of node2vec has limited capabilities to adapt to different learning rates. As a result, for boosting up the performance of the link prediction task, we built the deep learning model with adaptive learning rate optimizers: Adam, Adamax, Adadelta, and Adagrad, respectively. This approach is treated as a supervised classification problem, where the network aims to yield a single value representing the probability for a given edge. Thus, we end the deep learning model with a sigmoid activation function to score between 0 and 1.

\subsubsection{Optimizers.}
Below we have discussed the different types of optimizers that were used with the model.
\begin{itemize}
  \item Adagrad: Adaptive gradient, or AdaGrad, divides the learning rate by the square root of $v$, which is mainly the cumulative sum of current and past squared gradients up to time $t$ \cite{duchi2011adaptive}. Moreover, the gradient component is unchanged just like in SGD. The Adagrad is defined as such:
  \begin{equation}\label{twentyone_equ}
    w_{t+1} = w_t-\frac{\rho}{\sqrt{v_t+\epsilon}}\cdot \frac{\partial L}{\partial w_t}
    \end{equation}
    
     In Equation (\ref{twentyone_equ}), $w_t$ is the current weight at time step $t$ that needs to be updated, $\rho$ represents the learning rate and $\frac{\partial L}{\partial w_t}$ denotes the gradient descent to update the weight at $w_t$ and $\epsilon$ is a constant value.
    \vspace{3mm}
    \item Adadelta: Adadelta is a much more powerful extension of Adagrad that emphasizes the learning rate component \cite{zeiler2012adadelta}. The optimizer is based on updating gradient using the sliding window technique instead of aggregating all the previous gradients. In Adadelta, the difference between the current and updated weights is denoted by `delta'. Furthermore, the learning rate parameter is replaced by $T$, the exponential moving average of squared deltas and is defined in \ref{twentytwo_equ}).
    \begin{equation}\label{twentytwo_equ}
    w_{t+1} = w_t-\frac{\sqrt{T_{t-1}+\epsilon}}{\sqrt{v_t+\epsilon}}\cdot \frac{\partial L}{\partial w_t}
    \end{equation}
    
  \item Adam: Adaptive moment estimation, or Adam, focuses on the gradient component by using $\hat s$, which estimates the exponential average of the moving gradients \cite{kingma2014adam}. In addition, the learning rate component is calculated by dividing the learning rate $\rho$ by the square root of $v$, which is the exponential moving average of squared gradients. The equation is defined as below:
      \begin{equation}\label{twentythr_equ}
       w_{t+1} = w_t-\frac{\sqrt{T_{t-1}+\epsilon}}{\sqrt{v_t}+\epsilon}\cdot \hat s_t 
    \end{equation}
    \item Adamax: AdaMax is a variation of the Adam optimizer, which uses infinity norms \cite{kingma2014adam}. The infinity norm is used to calculate the absolute values of the $v$ components in a vector space (`max'), and $\hat s$ refers to the estimated value of the exponential average of moving gradients, and $v$ is the exponential moving average of previous $p$-norm of gradients, that is approximately the max function as defined below:
  \begin{equation}\label{twentyf_equ}
       w_{t+1} = w_t-\frac{\rho}{v_t}\cdot \hat s_t 
    \end{equation}
\end{itemize}

\section{Experiment}\label{Experiments}

This section will evaluate the proposed model on real-world data network datasets and examine how it is more effective than the existing benchmark methods, including Adamic Adar, Preferential Attachment, and Jaccard Coefficient.

\subsection{Datasets}
We evaluate our model on Facebook\footnote{https://snap.stanford.edu/data/egonets-Facebook.html} and Twitter\footnote{https://snap.stanford.edu/data/egonets-Twitter.html} datasets that consists of nodes and edges. Table \ref{table:66} shows the overview of the five network datasets. The first four of these datasets were collected from the SNAP website. Also, with the aid of Twitter API we have extended a Twitter dataset of around 7,000 users who have followed Twitter medical accounts \cite{zainab}. The extended dataset contains the follower and following IDs of the users working in the medical profession. Public biographical contents of the users were used for finding the occupation of the users. \footnote{\href{https://github.com/ZainabKazi22/occupation\textunderscore twitter}{https://github.com/ZainabKazi22/occupation\textunderscore twitter}}.

\vspace{-4mm}

\begin{table}[]
\caption{Details of the datasets}
\begin{center}
\begin{tabular}{|l|l|l|}
\hline
\textbf{Dataset} & \textbf{Number of nodes} & \textbf{Number of edges} \\ \hline
Twitter          & 81,306                   & 1,768,149                \\ \hline
Facebook1        & 4,039                    & 88,234                   \\ \hline
Facebook2        & 1,046                    & 27,794                   \\ \hline
Facebook3        & 546                      & 5,360                    \\ \hline
Occupation       & 6,754                    & 470,168                  \\ \hline
\end{tabular}
\label{table:66}
\end{center}
\end{table}

\vspace{-10mm}

\subsection{Experimental Results \& Discussions}
Our proposed model was implemented in Python 2.8.6, and the experiment was conducted on HPC (High Configuration GPU enabled PC)\footnote{\href{https://github.com/ZainabKazi22/link\_prediction\_with\_gbm}{https://github.com/ZainabKazi22/link\_prediction\_with\_gbm}}. In our model, we have used four layer fully connected deep neural network with 1024 ReLU neurons in each of the hidden layers. Then, we developed our model using Adagrad, Adadelta, Adam, and Adamax optimizers to improve the performance of the link prediction task.

We calculate the Area Under ROC Curve (AUC) scores to evaluate the performance of our approach of combining node2vec and deep learning model with each of the optimizers, respectively. The AUC score is defined in (\ref{tf_equ}):
\begin{equation}\label{tf_equ}
AUC = \frac{D_0-n_0(n_0 +1)/2}{n_0n_1}
\end{equation}

In Equation (\ref{tf_equ}), $n_0$ and $n_1$ denotes the number of positive and negative class links, respectively and $ D_0 = \sum r_i $, where $r_i$ represents the rank of the index $i$ in the positive class link in terms of similarity index. Also, $AUC \in [0,1]$, in which the higher the value of $AUC$, the higher the link prediction accuracy of the algorithm. We have compared the performance of our approach with the traditional link prediction benchmark methods: Adamic Adar (AA), Jaccard Co-efficient (JC), and Preferential Attachment (PA). 
\begin{table*}[t]
\caption{AUC Scores of the Link Prediction Algorithms}
\begin{center}
\begin{adjustbox}{width=.9\textwidth,center}
\begin{tabular}{|l|lllll|}
\hline
{\textbf{Link Prediction Algorithm}} & \multicolumn{5}{c|}{\textbf{AUC Score}}             \\ \cline{2-6} 
                                                    & \multicolumn{1}{l|}{\textbf{Twitter}} & \multicolumn{1}{l|}{\textbf{Occupation}} & \multicolumn{1}{l|}{\textbf{Facebook1}} & \multicolumn{1}{l|}{\textbf{Facebook2}} & \textbf{Facebook3} \\ \hline
Node2vec                                            & \multicolumn{1}{l|}{0.895}            & \multicolumn{1}{l|}{0.876}               & \multicolumn{1}{l|}{0.938}              & \multicolumn{1}{l|}{0.873}              & 0.861              \\ \hline
Node2vec + DL (Adam)                                & \multicolumn{1}{l|}{0.902}            & \multicolumn{1}{l|}{0.931}               & \multicolumn{1}{l|}{0.934}              & \multicolumn{1}{l|}{0.862}              & 0.855              \\ \hline
Node2vec + DL (Adamax)                              & \multicolumn{1}{l|}{0.916}            & \multicolumn{1}{l|}{\textbf{0.945}}      & \multicolumn{1}{l|}{\textbf{0.941}}     & \multicolumn{1}{l|}{\textbf{0.879}}     & 0.882              \\ \hline
Node2vec + DL (Adagrad)                             & \multicolumn{1}{l|}{0.911}            & \multicolumn{1}{l|}{0.911}               & \multicolumn{1}{l|}{0.932}              & \multicolumn{1}{l|}{0.845}              & 0.851              \\ \hline
Node2vec + DL (Adadelta)                            & \multicolumn{1}{l|}{\textbf{0.924}}   & \multicolumn{1}{l|}{0.932}               & \multicolumn{1}{l|}{0.908}              & \multicolumn{1}{l|}{0.871}              & \textbf{0.863}     \\ \hline
Adamic Adar                                         & \multicolumn{1}{l|}{0.897}            & \multicolumn{1}{l|}{0.711}               & \multicolumn{1}{l|}{0.898}              & \multicolumn{1}{l|}{0.878}              & 0.734              \\ \hline
Jaccard Co-efficient                                & \multicolumn{1}{l|}{0.897}            & \multicolumn{1}{l|}{0.748}               & \multicolumn{1}{l|}{0.901}              & \multicolumn{1}{l|}{0.856}              & 0.699              \\ \hline
Preferential Attachment                             & \multicolumn{1}{l|}{0.891}            & \multicolumn{1}{l|}{0.803}               & \multicolumn{1}{l|}{0.835}              & \multicolumn{1}{l|}{0.801}              & 0.76               \\ \hline
\end{tabular}%
\end{adjustbox}
\end{center}
\label{table:77}
\end{table*}

In AA, the association between two neighboring nodes with a smaller degree may occur more than a node with a higher degree \cite{zhou2009predicting}. For instance, two celebrity fans are likely not to know each other. Yet, if two users follow someone who has fewer fans, then those two users will have a higher chance to have similar interests or tend to be in the same social circle. JC believes that the probability of the presence of links is proportional to the number of two nodes' neighbors \cite{gupta2015significance}.  If any two Twitter users tend to have similar interests, they have a higher chance of having some type of connections with each other. Research has shown that the rate of an edge to be connected to a node is proportional to the degree of the node \cite{abbasi2012betweenness}. Thus, PA states that the chances of a new edge to be connected to a node are related to the degree of the node. Two popular celebrities will have a higher chance to know each other since they have a higher degree compared to two ordinary persons. The equations of the following algorithms are stated in (\ref{8}), (\ref{9}), and (\ref{10}):

\begin{equation}
s_{AA}= \sum \limits _{x\in \Gamma (i)\cap \Gamma (j)} {\frac {1} {\log k_{x}}}
\label{8}
\end{equation} 
\begin{equation}
s_{JC}= \frac {\left |{ {\Gamma (i)\cap \Gamma (j)} }\right |}{\left |{ {\Gamma (i)\cup \Gamma (j)} }\right |}
\label{9}
\end{equation}
\begin{equation}
s_{PA}=  k_{x} \cdot k_{y}
\label{10}
\end{equation}

Table \ref{table:77} shows the AUC scores obtained from the link prediction algorithms. Overall, node2vec and the node2vec optimized algorithm (Node2Vec+DL) have performed better than the traditional benchmark methods. This might be because node2vec algorithms can learn high-level features from the network data \cite{grover2016node2vec}. Moreover, as high-end robust computational engines like GPU are readily available, it is possible to execute the deep learning models. Whereas predicting future links from large network data is challenging for the existing benchmark methods. Node2vec with Adamax optimizer has received the highest AUC score in Occupation and Facebook1 and Facebook2 datasets among the deep learning models. The node2vec with Adadelta optimizer has performed best in Twitter and Facebook3 datasets. The model with Adamax optimizer has performed better than the rest of the optimizers across three datasets, proving that the Adamax optimizer modified over Adam optimizer performs better than the Adam optimizer. Similarly, the model with the Adadelta optimizer has performed better for Twitter and Facebook3 datasets than the Adagrad optimizer. This has demonstrated that the Adadelta optimizer, an improved version of Adagrad optimizer, has achieved a better performance score than the Adagrad optimizer. Thus, from the results in Table \ref{table:77}, we can see that optimizers of the DL model have increased the performance of the node2vec algorithm. The model proposed in this paper has acquired higher AUC scores than the existing benchmark and node2vec method. Also, the AUC scores of the node2vec with improved optimizers of the DL model are highest across all the datasets.

\section{Conclusion}\label{Conclusion}
In this paper, we explored the drawbacks of the node2vec algorithm when boosting up non-convex functions. In other words, the likelihood of falling into a local minimum due to lack of network knowledge and SGD optimizer's incapabilities to execute adaptive adjustment of the learning rate. Hence, such a scenario makes it extremely difficult for node2vec to process sparse social networks. As a result, we proposed NODDLE, a deep learning model where we have merged the features aggregated by the node2vec algorithm and used them as inputs into a multi-layer neural network optimizing its performance by using different types of improved optimizers such as  Adam, Adamax, Adadelta, and Adagrad. Compared to the various baselines, the results of experiments on real-world social networks proved that our approach enhances the prediction accuracy and is much more effective and efficient.

\subsubsection*{Acknowledgements.}

The authors thank DaTALab members \& Lakehead University's HPC (High Configuration GPU enabled PC) for executing the models, and Punardeep Sikkha, Arunim Garg, Bart, and Abhijit Rao for proofreading and reviewing the manuscript.

%
%
%
%
\bibliographystyle{splncs04}
\bibliography{refs}

\begin{thebibliography}{10}
\providecommand{\url}[1]{\texttt{#1}}
\providecommand{\urlprefix}{URL }
\providecommand{\doi}[1]{https://doi.org/#1}

\bibitem{abbasi2012betweenness}
Abbasi, A., Hossain, L., Leydesdorff, L.: Betweenness centrality as a driver of
  preferential attachment in the evolution of research collaboration networks.
  Journal of Informetrics  \textbf{6}(3),  403--412 (2012)

\bibitem{adamic2003friends}
Adamic, L.A., Adar, E.: Friends and neighbors on the web. Social networks
  \textbf{25}(3),  211--230 (2003)

\bibitem{aiello2012friendship}
Aiello, L.M., Barrat, A., Schifanella, R., Cattuto, C., Markines, B., Menczer,
  F.: Friendship prediction and homophily in social media. ACM Transactions on
  the Web (TWEB)  \textbf{6}(2),  1--33 (2012)

\bibitem{airoldi2008mixed}
Airoldi, E.M., Blei, D.M., Fienberg, S.E., Xing, E.P.: Mixed membership
  stochastic blockmodels. Journal of machine learning research
  \textbf{9}(Sep),  1981--2014 (2008)

\bibitem{al2006link}
Al~Hasan, M., Chaoji, V., Salem, S., Zaki, M.: Link prediction using supervised
  learning. In: SDM06: workshop on link analysis, counter-terrorism and
  security. vol.~30, pp. 798--805 (2006)

\bibitem{al2019ddgk}
Al-Rfou, R., Perozzi, B., Zelle, D.: Ddgk: Learning graph representations for
  deep divergence graph kernels. In: The World Wide Web Conference. pp. 37--48
  (2019)

\bibitem{amari1993backpropagation}
Amari, S.i.: Backpropagation and stochastic gradient descent method.
  Neurocomputing  \textbf{5}(4-5),  185--196 (1993)

\bibitem{ayoub2020accurate}
Ayoub, J., Lotfi, D., El~Marraki, M., Hammouch, A.: Accurate link prediction
  method based on path length between a pair of unlinked nodes and their
  degree. Social Network Analysis and Mining  \textbf{10}(1), ~9 (2020)

\bibitem{benchettara2010supervised}
Benchettara, N., Kanawati, R., Rouveirol, C.: Supervised machine learning
  applied to link prediction in bipartite social networks. In: 2010
  International Conference on Advances in Social Networks Analysis and Mining.
  pp. 326--330. IEEE (2010)

\bibitem{bressan2017counting}
Bressan, M., Chierichetti, F., Kumar, R., Leucci, S., Panconesi, A.: Counting
  graphlets: Space vs time. In: Proceedings of the Tenth ACM International
  Conference on Web Search and Data Mining. pp. 557--566 (2017)

\bibitem{canziani2016analysis}
Canziani, A., Paszke, A., Culurciello, E.: An analysis of deep neural network
  models for practical applications. arXiv preprint arXiv:1605.07678  (2016)

\bibitem{chen2017harp}
Chen, H., Perozzi, B., Hu, Y., Skiena, S.: Harp: Hierarchical representation
  learning for networks. arXiv preprint arXiv:1706.07845  (2017)

\bibitem{chen2018harp}
Chen, H., Perozzi, B., Hu, Y., Skiena, S.: Harp: Hierarchical representation
  learning for networks. In: Proceedings of the AAAI Conference on Artificial
  Intelligence. vol.~32 (2018)

\bibitem{chen2019n2vscdnnr}
Chen, J., Wu, Y., Fan, L., Lin, X., Zheng, H., Yu, S., Xuan, Q.: N2vscdnnr: A
  local recommender system based on node2vec and rich information network. IEEE
  Transactions on Computational Social Systems  \textbf{6}(3),  456--466 (2019)

\bibitem{deylami2015link}
Deylami, H.A., Asadpour, M.: Link prediction in social networks using
  hierarchical community detection. In: 2015 7th Conference on Information and
  Knowledge Technology (IKT). pp.~1--5. IEEE (2015)

\bibitem{doppa2010learning}
Doppa, J.R., Yu, J., Tadepalli, P., Getoor, L.: Learning algorithms for link
  prediction based on chance constraints. In: Joint european conference on
  machine learning and knowledge discovery in databases. pp. 344--360. Springer
  (2010)

\bibitem{duchi2011adaptive}
Duchi, J., Hazan, E., Singer, Y.: Adaptive subgradient methods for online
  learning and stochastic optimization. Journal of machine learning research
  \textbf{12}(7) (2011)

\bibitem{goldberg2014word2vec}
Goldberg, Y., Levy, O.: word2vec explained: deriving mikolov et al.'s
  negative-sampling word-embedding method. arXiv preprint arXiv:1402.3722
  (2014)

\bibitem{grover2016node2vec}
Grover, A., Leskovec, J.: node2vec: Scalable feature learning for networks. In:
  Proceedings of the 22nd ACM SIGKDD international conference on Knowledge
  discovery and data mining. pp. 855--864 (2016)

\bibitem{gupta2015significance}
Gupta, A.K., Sardana, N.: Significance of clustering coefficient over jaccard
  index. In: 2015 Eighth International Conference on Contemporary Computing
  (IC3). pp. 463--466. IEEE (2015)

\bibitem{hu2017retracted}
Hu, W., Wang, H., Peng, C., Liang, H., Du, B.: Retracted: An event detection
  method for social networks based on link prediction (2017)

\bibitem{hu2017event}
Hu, W., Wang, H., Qiu, Z., Nie, C., Yan, L., Du, B.: An event detection method
  for social networks based on hybrid link prediction and quantum swarm
  intelligent. World Wide Web  \textbf{20}(4),  775--795 (2017)

\bibitem{jain2017non}
Jain, P., Kar, P.: Non-convex optimization for machine learning. arXiv preprint
  arXiv:1712.07897  (2017)

\bibitem{katz1953new}
Katz, L.: A new status index derived from sociometric analysis. Psychometrika
  \textbf{18}(1),  39--43 (1953)

\bibitem{kaya2017unsupervised}
Kaya, M., Jawed, M., B{\"u}t{\"u}n, E., Alhajj, R.: Unsupervised link
  prediction based on time frames in weighted--directed citation networks. In:
  Trends in Social Network Analysis, pp. 189--205. Springer (2017)

\bibitem{zainab}
Khanam, K.Z., Srivastava, G., Mago, V.: Identifying health related occupations
  of twitter users through word embedding and deep neural networks. Proceedings
  of The 19th Asia Pacific Bioinformatics Conference  \textbf{Accepted, In
  press} (2021)

\bibitem{kingma2014adam}
Kingma, D.P., Ba, J.: Adam: A method for stochastic optimization. arXiv
  preprint arXiv:1412.6980  (2014)

\bibitem{le2014distributed}
Le, Q., Mikolov, T.: Distributed representations of sentences and documents.
  In: International conference on machine learning. pp. 1188--1196 (2014)

\bibitem{li2018link}
Li, J.c., Zhao, D.l., Ge, B.F., Yang, K.W., Chen, Y.W.: A link prediction
  method for heterogeneous networks based on bp neural network. Physica A:
  Statistical Mechanics and its Applications  \textbf{495},  1--17 (2018)

\bibitem{li2018deep}
Li, T., Zhang, J., Philip, S.Y., Zhang, Y., Yan, Y.: Deep dynamic network
  embedding for link prediction. IEEE Access  \textbf{6},  29219--29230 (2018)

\bibitem{li2014deep}
Li, X., Du, N., Li, H., Li, K., Gao, J., Zhang, A.: A deep learning approach to
  link prediction in dynamic networks. In: Proceedings of the 2014 SIAM
  International Conference on Data Mining. pp. 289--297. SIAM (2014)

\bibitem{liben2007link}
Liben-Nowell, D., Kleinberg, J.: The link-prediction problem for social
  networks. Journal of the American society for information science and
  technology  \textbf{58}(7),  1019--1031 (2007)

\bibitem{liu2020network}
Liu, D., Li, Q., Ru, Y., Zhang, J.: The network representation learning
  algorithm based on semi-supervised random walk. IEEE Access  (2020)

\bibitem{liu2019link}
Liu, H., Kou, H., Yan, C., Qi, L.: Link prediction in paper citation network to
  construct paper correlation graph. EURASIP Journal on Wireless Communications
  and Networking  \textbf{2019}(1),  1--12 (2019)

\bibitem{maron2019provably}
Maron, H., Ben-Hamu, H., Serviansky, H., Lipman, Y.: Provably powerful graph
  networks. In: Advances in neural information processing systems. pp.
  2156--2167 (2019)

\bibitem{ozcan2018link}
Ozcan, A., Oguducu, S.G.: Link prediction in evolving heterogeneous networks
  using the narx neural networks. Knowledge and Information Systems
  \textbf{55}(2),  333--360 (2018)

\bibitem{pezeshkpour2019investigating}
Pezeshkpour, P., Tian, Y., Singh, S.: Investigating robustness and
  interpretability of link prediction via adversarial modifications. arXiv
  preprint arXiv:1905.00563  (2019)

\bibitem{rahman2016link}
Rahman, M., Al~Hasan, M.: Link prediction in dynamic networks using graphlet.
  In: Joint European Conference on Machine Learning and Knowledge Discovery in
  Databases. pp. 394--409. Springer (2016)

\bibitem{rahman2018dylink2vec}
Rahman, M., Saha, T.K., Hasan, M.A., Xu, K.S., Reddy, C.K.: Dylink2vec:
  Effective feature representation for link prediction in dynamic networks.
  arXiv preprint arXiv:1804.05755  (2018)

\bibitem{ruder2016overview}
Ruder, S.: An overview of gradient descent optimization algorithms. arXiv
  preprint arXiv:1609.04747  (2016)

\bibitem{tabakhi2014unsupervised}
Tabakhi, S., Moradi, P., Akhlaghian, F.: An unsupervised feature selection
  algorithm based on ant colony optimization. Engineering Applications of
  Artificial Intelligence  \textbf{32},  112--123 (2014)

\bibitem{tang2015line}
Tang, J., Qu, M., Wang, M., Zhang, M., Yan, J., Mei, Q.: Line: Large-scale
  information network embedding. In: Proceedings of the 24th international
  conference on world wide web. pp. 1067--1077 (2015)

\bibitem{wang2017relational}
Wang, H., Shi, X., Yeung, D.Y.: Relational deep learning: A deep latent
  variable model for link prediction. In: AAAI. pp. 2688--2694 (2017)

\bibitem{wang2015link}
Wang, P., Xu, B., Wu, Y., Zhou, X.: Link prediction in social networks: the
  state-of-the-art. Science China Information Sciences  \textbf{58}(1),  1--38
  (2015)

\bibitem{xu2018interaction}
Xu, L., Wei, X., Cao, J., Philip, S.Y.: Interaction content aware network
  embedding via co-embedding of nodes and edges. In: Pacific-Asia Conference on
  Knowledge Discovery and Data Mining. pp. 183--195. Springer (2018)

\bibitem{yao2016link}
Yao, L., Wang, L., Pan, L., Yao, K.: Link prediction based on common-neighbors
  for dynamic social network. Procedia Computer Science  \textbf{83},  82--89
  (2016)

\bibitem{zeiler2012adadelta}
Zeiler, M.D.: Adadelta: an adaptive learning rate method. arXiv preprint
  arXiv:1212.5701  (2012)

\bibitem{zhang2016deep}
Zhang, C., Zhang, H., Yuan, D., Zhang, M.: Deep learning based link prediction
  with social pattern and external attribute knowledge in bibliographic
  networks. In: 2016 IEEE International Conference on Internet of Things
  (iThings) and IEEE Green Computing and Communications (GreenCom) and IEEE
  Cyber, Physical and Social Computing (CPSCom) and IEEE Smart Data
  (SmartData). pp. 815--821. IEEE (2016)

\bibitem{zhang2016analytical}
Zhang, L., Zhuo, F., Bai, C., Xu, H.: Analytical model for predictable contact
  in intermittently connected cognitive radio ad hoc networks. International
  Journal of Distributed Sensor Networks  \textbf{12}(7),  1550147716659426
  (2016)

\bibitem{zhang2017efficient}
Zhang, Z., Wen, J., Sun, L., Deng, Q., Su, S., Yao, P.: Efficient incremental
  dynamic link prediction algorithms in social network. Knowledge-Based Systems
   \textbf{132},  226--235 (2017)

\bibitem{zhou2009predicting}
Zhou, T., L{\"u}, L., Zhang, Y.C.: Predicting missing links via local
  information. The European Physical Journal B  \textbf{71}(4),  623--630
  (2009)

\bibitem{zhu2020semi}
Zhu, J., Zheng, Z., Yang, M., Fung, G.P.C., Tang, Y.: A semi-supervised model
  for knowledge graph embedding. Data Mining and Knowledge Discovery
  \textbf{34}(1),  1--20 (2020)

\end{thebibliography}
\end{document}